\documentclass{PoS}

\newcommand{\MeV}{{\rm \,MeV}}

\newcommand{\HQSS}{{\rm HQSS}}

\newcommand{\SU}{\mbox{SU}}
\newcommand{\U}{\mbox{U}}
\newcommand{\bc}{{\bar{c}}}
\newcommand{\bq}{{\bar{q}}}

\newcommand{\ben}{\begin{enumerate}}
\newcommand{\een}{\end{enumerate}}

\newcommand{\be}{\begin{equation}}
\newcommand{\ee}{\end{equation}}
\newcommand{\bea}{\begin{eqnarray}}
\newcommand{\eea}{\end{eqnarray}}
\newcommand{\ds}{\begin{displaystyle}}
\renewcommand{\ss}{\begin{scriptstyle}}

\newcommand{\ignore}[1]{}

\newcommand{\ba}{\begin{eqnarray}}
\newcommand{\ea}{\end{eqnarray}}

\title{Dynamically-generated baryon resonances with heavy flavor}

\ShortTitle{Dynamically-generated baryon resonances with heavy flavor}

\author{\speaker{Olena Romanets}%
         \thanks{Supported by Spanish Ministerio de Econom{\'\i}a y Competitividad
(FIS2011-28853-C02-02, FIS2011-24149, FPA2010-16963),
Junta de Andalucia (FQM-225), Generalitat Valenciana (PROMETEO/2009/0090) and EU HadronPhysics2 project (grant 227431). O. R. acknowledges support from the Rosalind Franklin Fellowship.
L. T. acknowledges support from RyC Program, and FP7-PEOPLE-2011-CIG (PCIG09-GA-2011-291679).}\\
        KVI, University of Groningen, Zernikelaan~25,
    9747~AA~Groningen, The~Netherlands\\
        E-mail: \email{olena.romanets@gmail.com}}

\author{Carmen Garc\'ia-Recio and Lorenzo Luis Salcedo\\
        Departamento~de~F{\'\i}sica~At\'omica, Molecular~y~Nuclear,
 and Instituto
  Carlos I de F{\'\i}sica Te\'orica y Computacional, Universidad~de~Granada,
Granada,  E-18071, Spain\\
}
\author{Juan Nieves\\
Instituto~de~F{\'\i}sica~Corpuscular~(centro~mixto~CSIC-UV),
  Institutos~de~Investigaci\'on~de~Paterna,
 Aptdo.~22085,~Valencia,~46071,~Spain
}

\author{Laura Tolos\\
Institut~de~Ci\`encies~de~l'Espai~(IEEC/CSIC), 
  Campus~Universitat~Aut\`onoma~de~Barcelona, Facultat~de~Ci\`encies,
  Torre~C5,~Bellaterra,~E-08193,~Spain
}

\abstract{
We study baryon resonances with heavy flavor in a molecular approach, 
thus as dynamically generated by meson-baryon scattering. This is accomplished 
by using a unitary coupled-channel model taking, as bare interaction, the extension 
to four flavors and spin symmetry of the Weinberg-Tomozawa interaction potential. 
A special attention is paid to the inclusion of heavy-quark spin symmetry and to 
the studies of the generated baryon resonances which complete the heavy-quark spin multiplets.
We reproduce few charmed and strange baryon resonances found experimentally, 
that is, $\Lambda_c(2595)$, $\Lambda_c(2625)$, $\Xi_c(2790)$, $\Xi_c(2815)$, and make predictions 
for more states. The bottom-flavored 
$\Lambda_b(5912)$ and $\Lambda_b(5920)$ states, found by the LHCb collaboration, 
are also obtained by our model, thus these resonances can be interpreted as molecular states. 
We also study $\Xi_b$ resonances, which belong to the same ${\rm SU(3)} \times {\rm HQSS}$ multiplets as  the 
observed $\Lambda_b$ particles. Finally we analyze hidden-charm baryon resonances. 
In this sector, we predict seven $N$-like and five $\Delta$-like states with masses around 4~GeV, 
most of them as bound states, and compare them with predictions from other models.
The predicted states can be searched for in future experiments that involve studies of heavy-flavor  
physics, e.g. PANDA/FAIR, where charm physics will be analyzed.
}

\FullConference{XV International Conference on Hadron Spectroscopy \\
                 4-8/11/2013\\
                 Nara, Japan}

\begin{document}

\section{Introduction}
The study of particles with heavy flavor is an active research topic,
and among them baryon resonances have recently attracted a lot of attention.
An interest in heavy-flavor physics in the last years was induced by the
 past and 
on-going experiments such as CLEO, Belle, {\it BABAR}, LHCb~\cite{experiments1}.
Moreover,
the planned PANDA and CBM experiments at FAIR~\cite{experiments2} will bring even more 
excitement in the field.
It is an important task to understand the nature of possible new states,
e.g. whether baryon resonances can be interpreted as three-quark states, or better as 
molecular states.

In this work we study baryon resonances with heavy (charm and bottom) flavor as molecular states,
thus dynamically generated by the meson-baryon dynamics.
There are several approaches used to study this type of states.
The coupled-channels models turned out to be successful in describing some 
of the experimental data,
and include unitarized coupled-channels
models
\cite{unitarizedcchm, TVME, JimenezTejero:2009vq, HofmannLutz},
the J\"ulich meson-exchange model~\cite{Julich}
and schemes based on hidden gauge formalism~\cite{Wu:2010jy}.
In this work we review findings of the unitarized coupled-channels model based on the 
spin-flavor extension of the Weinberg-Tomozawa (WT) potential, that has been used for
studying charmed strange and non-strange baryon resonances~\cite{PaperOpen},
hidden-charm baryon resonances~\cite{PaperHidden}, 
and bottom states~\cite{PaperBottom}.

\section{The phenomenological model}
We use an extension of the WT contact term interaction to $\rm SU(8)$ 
spin-flavor symmetry. 
The extended WT Hamiltonian  for four flavors and three colors
reads~\cite{GarciaRecio:2006wb}
\begin{equation}
{H}_{\rm WT}^{\rm sf}(x)
= -\frac{{\rm i}}{4f^2} :[\Phi, \partial_0 \Phi]^A{}_B
{\cal B}^\dagger_{ACD} {\cal B}^{BCD}:
,
\quad
A,B,\ldots = 1,\ldots,8
,
\label{eq:2.9}
\end{equation}
where $\Phi^A{}_B(x)$ is the meson field,  
which contains the fields of $0^-$ (pseudoscalar) and $1^-$ (vector)
mesons,
and
${\cal B}^{ABC}$ is the baryon field, which is a completely symmetric
tensor, containing the lowest-lying baryons
with $J^P=\frac{1}{2}^+$ and $\frac{3}{2}^+$.
The Hamiltonian can be rewritten as
\begin{equation}
{H}_{\rm WT} = H_{\rm ex} + H_{\rm ac},
\end{equation}
where $H_{\rm ex}$ is the exchange part, in which a quark is
transferred from the meson to the baryon, and another one is transferred from
the baryon to the meson, and $H_{\rm ac}$ is the annihilation-creation  mechanism, where
an antiquark in the meson annihilates with a similar quark
in the baryon, with subsequent creation of a quark and an antiquark.
It turns out~\cite{PaperHidden} that
the annihilation-creation part of the Hamiltonian can violate the heavy-quark spin symmetry
(HQSS)
 when the annihilation or
creation of $q\bq$ pairs involves heavy flavor,
since
according to the HQSS the number of heavy quarks and the number of heavy 
antiquarks are separately conserved (this implies $\U_c(1)\times \U_\bc(1)$).
The HQSS group, which also includes a group of separate rotations of the $c$ quark and 
$\bar c$ antiquark, reads as
$\SU_c(2) \times \SU_\bc(2) \times \U_c(1)\times \U_\bc(1)$.
We modify the Hamiltonian in correspondence to the requirements of HQSS.
Thus, in the sector of open-charm and open-bottom baryon resonances we 
eliminate possible coupled channels that contain heavy quark-heavy antiquark, since in this case 
the annihilation-creation mechanism would necessarily contain the heavy flavor.
In the hidden-charm sectors, where in $H_{\rm ac}$ the annihilated and created antiquark 
is necessarily $\bar c$, the annihilation-creation part of the Hamiltonian is not present due to HQSS.~\cite{PaperHidden}.
After implementing this change into the Hamiltonian, the $\SU(8)$ symmetry is broken and the model possess
$\SU(6) \times \HQSS$ symmetry, where $\SU(6)$ is the spin-flavor symmetry for the three light flavors. 
 
We calculate the scattering amplitude by solving
the on-shell Bethe-Salpeter equation (BSE) in coupled channels.
%\begin{equation}
%T^{CSIJ}=(1-V^{CSIJ}G^{CSIJ})^{-1}V^{CSIJ}
%\label{eq:bse} \ ,
%\end{equation}
%where $V^{CSIJ}$ is the potential coming from the extended WT interaction,
%an $G^{CSIJ}$ is a diagonal matrix containing the baryon-meson propagator
%for each channel; 
The meson-baryon
propagators that appear in the BSE are regularized using the subtraction point method (see~\cite{PaperOpen} for details).
Baryon resonances appear as poles of the scattering amplitude
on the complex energy plane.
Poles on the first
Riemann sheet that appear on the real axis below threshold are 
bound states, and the ones in the second Riemann sheet below the
real axis and above threshold are identified with resonances.
%%\footnote{
%Often we
%  refer to all poles generically as resonances, regardless of their concrete
%  nature, since usually they can decay through other channels not included in
%  the model space.
%}.  
%
%
%The mass and the width of the state can be found from
%the position of the pole on the complex energy plane, and the 
%coupling of the resonance to different meson-baryon channels are found from
%the residues of the scattering amplitude close to the pole. 
We find the mass and the width as well as the couplings to the meson-baryon channels of 
a baryon resonance from the behavior of the scattering amplitude around the pole, that is, 
the coordinate of the pole in the complex energy plane and the residue, respectively. 

The $\SU(6) \times \HQSS$ symmetry
is broken in nature, thus, we break it adiabatically to the $\SU(2)$ isospin symmetry,
by implementing the physical hadron masses and meson decay constants~\cite{PaperOpen}.
In this way we could follow the evolution of the poles while breaking the symmetry and
classify the found states under the corresponding $\SU(6)$, $\SU(3)$, and HQSS group multiplets.

\section{Results for dynamically-generated baryon resonances}
\subsection{Open-charm baryon resonances}
With the described model a number of baryon resonances have been studied.
Let us present our results for the charmed baryon resonances with charm $C=1$.
Such states will be studied in the upcoming PANDA experiment at FAIR~\cite{experiments2}.

We have studied non-strange baryon resonances with $C=1$ and isospin $0$,
these are $\Lambda_c$ states. 
We find the experimentally known $\Lambda_c(2595)$ resonance. In our model it appears 
around $2618.8 - i 0.6\,\MeV$, and has a possible decay channel $\Sigma_c \pi$.
The experimental value of the width of $\Lambda_c(2595)$
$3.6^{+ 2.0}_{ -1.3}\,\MeV$ is not reproduced,
since we have not included the three-body decay
channel $\Lambda_c \pi \pi$, which represents about 67\% of the
decay events \cite{PDG}. 
We also find another $J=1/2$ baryon resonance with a very close mass of 2617.3~MeV,
but with a big width of about 90~MeV. This resonance together with
$\Lambda_c(2595)$ makes a two-pole structure, similar to the case of
$\Lambda(1405)$.
Moreover, this resonance forms a HQSS doublet
with the pole at $2666.6 - i 27\,\MeV$, which we identify with the
$\Lambda_c(2625)$ $J=3/2$ resonance, coupling strongly to the $\Sigma_c^* \pi$ channel. 
The experimental value of the width ($\Gamma < 1.9\,\MeV$) is not reproduced,
but if we lower the mass of the found resonance to 2625 MeV by
slightly changing the subtraction point,
 the width will decrease significantly.
There is another predicted  $\Lambda_c$ resonance, with the mass 2828.4~MeV and a small width. 
%It has a decay channel $N D$.

We predict the existence of three $J=1/2$ $\Sigma_c$ resonances with masses
2571.5, 2622.7, and 2643.4~MeV. The first one has a narrow width, and the other two
are broad ($\Gamma=188$ and $87\,\MeV$, respectively). Besides, our model
predicts two $J=3/2$ $\Sigma_c$ states, a bound state at 2568.4~MeV, and 
a broad ($\Gamma=67\,\MeV$) resonance at 2692.9~MeV. The verification 
of these predictions requires more experimental data.

Further, we find nine $\Xi_c$ baryon resonances. One of them we identify 
with the experimental $\Xi_c(2790)$ ($J=1/2$) resonance with the decay channel $\Xi'_c \pi$,
and another one with the experimental $\Xi_c(2815)$ ($J=3/2$), with the decay channel $\Xi^*_c \pi$.
The predicted features of the other seven $\Xi_c$ resonances can be found in Ref.~\cite{PaperOpen}.
We also predict the existence of five $\Omega_c$ states, all of them as bound states,
three with $J=1/2$, and two with 
$J=3/2$. The masses of these states are predicted to lie between 2800 and 3000~MeV.

\subsection{Hidden-charm baryon resonances}
We have studied $N$-like and $\Delta$-like baryon resonances 
with hidden charm, thus containing $c \bar c$.
We find seven $N$-like states, five of them as bound states, and
two of them with the small widths (0.1 and 2.8 MeV), with masses
between 3900 and 4050~MeV. Besides, we predict five $\Delta$-like
hidden-charm particles, all of them as bound states, with masses 
between 4000 and 4310~MeV. The two heaviest ones, with masses 
4306.2 and 4306.8~MeV, lie very close to the $\Delta J/\psi$
threshold (4306.9~MeV), and therefore they appear as cusps.
The general pattern of the hidden-charm states is that they couple weakly to $N \eta_c$ and $N J/\psi$
channels, and have significantly bigger couplings to the channels that contain $\Lambda_c, \Sigma_c$ or $\Sigma^*_c$ baryon,
and $\bar D$ or $\bar D^*$ meson.

The states in this sector form HQSS multiplets (not only doublets as it is in the open-charm sector).
%This degeneracy of the of the states with exact HQSS comes from the structure of the group multiplets.
%There are two $\SU(6) \times \HQSS$ multiples, with dimensions 56 and 70. 
Many of the states
differ from each other by the way the spin of the heavy quark-antiquark pair couples 
to the light-quarks content. Because HQSS implies that the interaction dependent 
on the spin state of the heavy quark vanishes in the infinite quark limit, such states are degenerate 
under HQSS.

The $N$-like crypto-exotic resonances were previously predicted by the 
zero-range vector exchange model~\cite{HofmannLutz}. 
The $N$ resonances studied in Ref.~\cite{HofmannLutz}
are about 500 MeV lighter than those found in our model.
%In contrast to our model, the zero-range vector exchange model 
%does not take into account vector mesons.
%
The hidden-gauge
formalism predicts the masses of hidden-charm $N$ to be about 400 MeV larger~\cite{Oset, XiaoNievesOset}.
It has been shown that 
this difference is mainly due to a different renormalization scheme~\cite{XiaoNievesOset}.
The masses of $N$-like states predicted by our model lie 
close to the ones predicted by the constituent 
quark model of Ref.~\cite{Yuan:2012wz} using the hyperfine chiral interaction based on 
meson exchange, whereas the instanton-induced
and color-magnetic hyperfine interactions produce higher masses for the resonances~\cite{Yuan:2012wz}.

\subsection{Baryon resonances with bottom}
Our model reproduces the $\Lambda_b(5912)$ and $\Lambda_b(5920)$
resonances, which were discovered by the LHCb collaboration in the $\Lambda_b \pi \pi$ spectrum~\cite{Aaij:2012da}.
In our scheme these states form a HQSS doublet, which
explains the closeness of their masses. 
We find $\Lambda_b(5912)$ and $\Lambda_b(5920)$ as bound states,
as we  
do not consider three-body decay channels.
% such as $\Lambda_b \pi \pi$.
Such decay can be realized 
through the intermediate decay to $\Sigma_b \pi$ [$\Sigma^*_b \pi$
for $\Lambda_b(5920)$], with consequential decay of the virtual $\Sigma_b$ ($\Sigma^*_b$)
to $\Lambda_b \pi$.

We also find a $\Lambda_b$ bound state with a mass 5797.6~MeV, without any possible strong decay,
and a bound state with a mass of 6009.3 MeV that decays to 
$\Sigma_b \pi$.
Apart from studying $\Lambda_b$ resonances, we have also made 
predictions for $\Xi_b$ resonances, which belong to the same $\SU(3)\times \HQSS$ irreps
as $\Lambda_b(5912)$ and $\Lambda_b(5920)$. We find three spin-1/2 $\Xi_b$ states,
with zero or small widths and masses 5874., 6035.4, and 6072.8~MeV, and one
spin-3/2 $\Xi_b$ bound state with a mass of 6043.3~MeV.

%There exists an old prediction by a relativized quark model of Ref.~\cite{Capstick:1986bm},
%where masses two orbitally-excited $\Lambda_b$ states
%are in good agreement with the experimental data. Thus, our model
%proposes an alternative explanation of the nature of the found $\Lambda_b(5912)$ 
%and $\Lambda_b(5920)$ resonances. 

\section{Summary}
A number of heavy-flavored baryon resonances, namely
charmed baryon resonances, hidden-charm states, and baryon resonances with bottom
have been studied using a coupled-channels unitary approach.
For this purpose we use a spin-flavor extension of the WT interaction to four flavors
implementing HQSS constraints. 
 We have obtained a number of $C=1$ baryon resonances, among them
experimentally confirmed $\Lambda_c(2595)$, $\Lambda_c(2625)$, $\Xi_c(2790)$,
and $\Xi_(2815)$, while the rest are predictions of the model. 
We have also studied hidden-charm $N$ and $\Delta$-like states, and compared 
them with the predictions of other theoretical models.
New experimental data is expected in the future PANDA experiment at FAIR, which will help 
to constrain our model as well as verify our predictions.
Finally, baryon resonances with bottom have been studied. Our model generates experimental
$\Lambda_b(5912)$ and $\Lambda_b(5920)$ states as molecular states, and we have predicted the existence
of other two $\Lambda_b$ resonances and few $\Xi_b$ states.


\begin{thebibliography}{99}

\bibitem{experiments1}
www.lepp.cornell.edu/Research/EPP/CLEO,{\enskip}belle.kek.jp,
www-public.slac.stanford.edu/babar, http://lhcb-public.web.cern.ch/lhcb-public/.

\bibitem{experiments2}
http://www.fair-center.eu/.


\bibitem{unitarizedcchm} 
M.~F.~M.~Lutz and E.~E.~Kolomeitsev, {\it Nucl.\ Phys.\  A {\bf \it 730}, 110 (2004)};
M.~F.~M.~Lutz and E.~E.~Kolomeitsev, {\it Nucl.\ Phys.\ A {755}, 29 (2005)}.

\bibitem{TVME} L. Tolos, J. Schaffner-Bielich and A. Mishra, {\it Phys.\ Rev.\  C { 70}, 025203 (2004)};
T.~Mizutani and A.~Ramos, {\it Phys.\ Rev.\ C {74}, 065201 (2006)}.

%\cite{JimenezTejero:2009vq}
\bibitem{JimenezTejero:2009vq} 
  C.~E.~Jimenez-Tejero, A.~Ramos and I.~Vidana,
  {\it Phys.\ Rev.\ C { 80}, 055206 (2009)}.

\bibitem{HofmannLutz} 
  J.~Hofmann and M.~F.~M.~Lutz,
 {\it  Nucl.\ Phys.\ A { 763}, 90 (2005)}; 
  {\it Nucl.\ Phys.\  A { 776}, 17 (2006)}.


\bibitem{Julich} J.~Haidenbauer {\it et al.}, {\it Eur.\ Phys.\ J.\  A { 33}, 107 (2007)};
J.~Haidenbauer {\it et al.}, {\it Eur.\ Phys.\ J.\  A { 37}, 55 (2008)}.
%J.~Haidenbauer, G.~Krein, U.~G.~Meissner and L.~Tolos, Eur. Phys. J. A {\bf 47}, 18 (2011).

\bibitem{Wu:2010jy} J.~-J.~Wu {\it et al.} {\it Phys.\ Rev.\ Lett.\  { 105}, 232001 (2010)};
 K.~P.~Khemchandani et al., {\it Phys.\ Rev.\ D { 83}, 114041 (2011)}.



\bibitem{PaperOpen}
 O.~Romanets, L.~Tolos, C.~Garcia-Recio, J.~Nieves, L.~L.~Salcedo, and R.G.E. Timmermans,
{\it Phys. Rev. D { 85}, 114032 (2012)}.

\bibitem{PaperHidden}
 C.~Garcia-Recio, J.~Nieves, O.~Romanets, L.~L.~Salcedo, and L.~Tolos,
 {\it  Phys.\ Rev.\ D { 87}, 074034 (2013)}.

\bibitem{PaperBottom}
 C.~Garcia-Recio, J.~Nieves, O.~Romanets, L.~L.~Salcedo, and L.~Tolos,
  {\it Phys.\ Rev.\ D { 87}, 034032 (2013)}.


% model

\bibitem{GarciaRecio:2006wb} 
  C.~Garcia-Recio, J.~Nieves and L.~L.~Salcedo,
 {\it  Phys.\ Rev.\ D { 74}, 036004 (2006)}.

\bibitem{PDG}
J. Beringer et al. {\it (Particle Data Group)}, 
{\it Phys. Rev. D { 86}, 010001 (2012)}.


\bibitem{Oset}
  J.~-J.~Wu, R.{\it et al.},
  {\it Phys.\ Rev.\ C { 84}, 015202 (2011)};
  J.~-J.~Wu {\it et al.},
  {\it Phys.\ Rev.\ C { 85}, 044002 (2012)}.


\bibitem{XiaoNievesOset}
C.~W.~Xiao, 
 J.~Nieves, E.~Oset,
%{\it et al.},
{\it Phys. Rev. D { 88} 056012 (2013)}.



\bibitem{Yuan:2012wz}
  S.~G.~Yuan
% K.~W.~Wei, J.~He, H.~S.~Xu and B.~S.~Zou,
{\it et al.}
 {\it  Eur.\ Phys.\ J.\ A { 48},61 (2012)}.


%\cite{Aaij:2012da}
\bibitem{Aaij:2012da}
  R.~Aaij {\it et al.} {\it  [LHCb Collaboration],
  %``Observation of excited Lambda_b0 baryons,''
  Phys.\ Rev.\ Lett.\  { 109} (2012) 172003}.
%  [arXiv:1205.3452 [hep-ex]].
  %%CITATION = ARXIV:1205.3452;%%


\end{thebibliography}
\end{document}